\documentclass[oldversion,structabstract]{aa} % for the letters
\usepackage{graphicx,subfigure}
\usepackage{pstricks}
\usepackage{txfonts}

  \usepackage{natbib}
  \bibpunct{(}{)}{;}{a}{}{,}

\usepackage{ulem}

\newcommand{\micron}{\ensuremath{\mu}m}
\newcommand{\ie}{\textsl{i.e.}}
\newcommand{\eg}{\textsl{e.g.}}

\usepackage{hyperref}
\begin{document}
   \title{Integral-field spectroscopy of
     (90482)~Orcus-Vanth\thanks{Based on observations
      collected at the
      European Southern Observatory
      Very Large Telescope
      (programs ID:
      \href{http://archive.eso.org/wdb/wdb/eso/eso_archive_main/query?prog_id=284.C-5044\%28A\%29&max_rows_returned=1000}
           {284.C-5044} \&
      \href{http://archive.eso.org/wdb/wdb/eso/eso_archive_main/query?prog_id=384.C-0877\%28A\%29&max_rows_returned=1000}
           {384.C-0877}).}
     }

   \subtitle{}

   \author{%
%    Beno\^{i}t
    Beno\^{i}t Carry\inst{1,2,3}
    \and
    Daniel Hestroffer\inst{4}
    \and
    Francesca E. DeMeo\inst{2,5}
    \and
    Audrey Thirouin\inst{6}
    \and
    J\'er\^{o}me Berthier\inst{4}
    \and
    Pedro Lacerda\inst{7}
    \and
    Bruno Sicardy\inst{2,8,9}
    \and
    Alain Doressoundiram\inst{2}
    \and
    Christophe Dumas\inst{10}
    \and
    David Farrelly\inst{11}
    \and
    Thomas G. M\"{u}ller\inst{12}
  }

  \offprints{Beno\^{i}t Carry: benoit.carry@esa.int}

  \institute{%
    European Space Astronomy Centre, ESA,
    P.O. Box 78,
    28691 Villanueva de la Ca\~{n}ada, Madrid, Spain\\
    \email{benoit.carry@esa.int}
    \and
    LESIA, Observatoire de Paris, CNRS, 
    5 place Jules Janssen,
    92190 Meudon, France
    \and
    Universit\'e Paris 7 Denis-Diderot,
    5 rue Thomas Mann,
    75205 Paris CEDEX, France
    \and
    IMCCE, Observatoire de Paris, UPMC, CNRS, 77 av. Denfert Rochereau
    75014 Paris, France
    \and
    Department of Earth, Atmospheric, and Planetary Sciences, MIT,
    77 Massachusetts Avenue, Cambridge, MA 02139, USA 
    \and
    Instituto de Astrof\'isica de Andaluc\'ia, CSIC, Apt 3004, 18080 Granada, Spain
    \and
    Queen's University, Belfast, County Antrim BT7 1NN, Ireland
    \and
    Universit\'e Pierre et Marie Curie, 4, Place Jussieu, 75252 Paris
    cedex 5, France
    \and
    Institut Universitaire de France, 103, Bld Saint Michel, 75005 Paris, France
    \and
    Alonso de C\'ordova 3107, Vitacura, Casilla 19001, Santiago de Chile, Chile
    \and
    Utah State University, 0300 Old Main Hill, Logan, UT 84322, USA
    \and
    Max-Planck-Institut f\"ur extraterrestrische Physik (MPE),
    Giessenbachstrasse, 85748 Garching, Germany 
  }

  \date{Received ; accepted}

% \abstract{}{}{}{}{} 
% 5 {} token are mandatory
 
  \abstract{}
  % aims heading (mandatory)
   {We seek to constrain the surface composition of the
     Trans-Neptunian Object (90482) Orcus and its small satellite
     Vanth, as well as their mass and density.}
  % methods heading (mandatory)
   {We acquired near-infrared spectra (1.4--2.4 $\mu$m)
     of (90482) Orcus and its companion Vanth using the
     adaptive-optics-fed
     integral-field spectrograph SINFONI mounted on Yepun/UT4 at the
     European Southern Observatory Very Large Telescope.
    We took advantage of a very favorable appulse (separation of
    only 4\arcsec) between Orcus and
    the UCAC2 29643541 star ($m_R = 11.6$) to use the  
    adaptive optics mode of SINFONI, allowing both components to be
    spatially resolved and Vanth colors to be 
    extracted independently from Orcus.
   }
  % results heading (mandatory)
   {The spectrum of Orcus we obtain has the highest signal-to-noise
     ratio to date, and we confirm the presence of H$_2$O ice in
     crystalline form, together with the presence of an absorption
     band at 2.2\,\micron.
     We set an upper limit of about 2\% for the presence of methane,
     and 5\% for ethane.
     Because the methane alone cannot account
     for the 2.2\,\micron~band, the presence of ammonia is suggested to
     the level of a couple of percent.
     The colors of Vanth are found slightly redder than those of
     Orcus, but the large measurement
     uncertainties forbid us from drawing conclusions on the
     origin of the pair (capture or co-formation).
     Finally, we reset the orbital phase of Vanth around Orcus, and
     confirm the orbital parameters derived by
     Brown et al. (2010, AJ 139).
   }{}

   \keywords{%
   Kuiper belt objects: individual: (90482) Orcus;
   Methods: observational;
   Techniques: high angular resolution;
   Techniques: imaging spectroscopy}   

   \maketitle
%
%________________________________________________________________

%------------------------------------------------%
%------------------------------------------------%
%---- TAG ---------------------------------------%
%------------------------------------------------%
%------------------------------------------------%
\section{Introduction\label{sec: intro}}
  \indent Binaries in the Solar System are of 
  high importance because
  they provide the most direct and precise way to derive the mass of
  minor planets
  \citep[see][]{2002-AsteroidsIII-2.2-Hilton}.
  Combined with volume estimates, their densities can be calculated, providing hints
  on their composition and interior
  \citep[\eg,][]{2002-AsteroidsIII-2.2-Merline,2002-AsteroidsIII-4.2-Britt}.
  Subsequently, they constrain the characteristics of the most pristine material
  of the solar system, and further our understanding of planetary system formation
  and dynamical evolution.
  In this valuable context, 
  the Trans-Neptunian Binary (TNB)
  Orcus/Vanth system is of
  particular interest for the following reasons.\\
  \indent 1. With an estimated albedo of $\sim$27\%
  \citep{2010-AA-518-Lim}, Orcus is
  among the brightest known 
  Trans-Neptunian Objects (TNOs); and has
  a diameter of about 850 kilometers.\\
  \indent 2. Near-infrared spectroscopy of Orcus has revealed a
  surface rich in water ice in crystalline form
  \citep{2004-AA-422-Fornasier,
    2005-AA-437-deBergh,
    2005-ApJ-627-Trujillo,
    2008-AA-479-Barucci,
    2010-AA-521-DeMeo}.
  Moreover,
  \citet{2005-ApJ-627-Trujillo},
  \citet{2008-AA-479-Barucci}, and
  \citet{2010-AA-520-Delsanti}
  detected a weak
  band around 2.2 $\mu$m that could
  be associated with either
  methane (CH$_4$) or ammonia (NH$_3$).
  {The long term stability of all ices are affected by high energy photon bombardment 
  (causing photodissociation and sputtering),
  micrometeorite impacts, radioactive decay, and sublimation.
  Both methane and ammonia are expected to be
  destroyed by solar irradiation on short timescales
  \citep{1998-PSS-46-Strazzulla, 
    2003-EMP-92-Cooper, 2003-ApJ-590-Cottin}.}
  Ammonia's presence, if confirmed, would thus require an active process
  to resupply the surface with ammonia, such as impact gardening or, more favorably,
  cryovolcanism 
  \citep[the ammonia lowers the melting temperature of water
  ice and hence favors such mechanism as highlighted
  by][]{2007-AJ-663-Cook}. { Bodies in the outer solar system 
  that have methane on their surface have retained their atmospheres
  which would also have important implications for its discovery
  on the surface of Orcus.}\\ 
  \indent 3. Recent radiometric measurements from ESA Herschel
  \citep[Key Program ``TNOs are Cool!'', see][]{2009-EMP-105-Muller}
  have refined the size estimate of Orcus
  to 850\,$\pm$\,90\,km
  \citep{2010-AA-518-Lim}.
  The diameter estimate will potentially be improved from the
  stellar occultations expected for upcoming years.
  Thus the improvement of Vanth's orbit 
  \citep[upon the solution by][]{2010-AJ-139-Brown}
  will help determine the bulk density of Orcus.\\
  \indent We present here new spectro-imaging data obtained in 2010
  that provide constraints on the composition of Orcus and the orbit
  of Vanth.
  We describe in Section~\ref{sec: obs} the observations,
  list in Section~\ref{sec: reduc} the data reduction and spectral
  extraction steps,
  present in Section~\ref{sec: spec} the analysis of the colours and
  spectra of Orcus and Vanth, and
  detail in Section~\ref{sec: astro} the
  orbit computation and stellar occultations prediction.

%------------------------------------------------%
%------------------------------------------------%
%---- TAG ---------------------------------------%
%------------------------------------------------%
%------------------------------------------------%
\section{Observations\label{sec: obs}}
  \indent The brightness contrast
  ($\Delta m_V \sim 2.6$) and small apparent angular separation
  ($\sim$0.2\arcsec) between Orcus
  and its satellite Vanth require the use of a high angular-resolution
  camera/spectrograph to spatially resolve the system. 
  This means observations have to be conducted
  in the visible from the Hubble Space Telescope
  \citep[\eg,][]{2010-AJ-139-Brown}, or in the near-infrared
  with ground-based telescopes equipped with adaptive optics (AO) modules.
  The latter is of high interest for cold objects like TNOs
  because many ices display strong absorption bands in the near-infrared
  \citep[see][]{2008-SSBN-3-Barucci}.\\
  \indent However, because adaptive-optics systems require a bright ($m_V
  \le 15$) reference source
  (a Natural Guide Star: NGS)
  to correct the incident wavefront from the
  deformations induced by the atmospheric turbulence, study of TNOs from the
  ground with AO is generally limited to the brightest objects
  (\eg, Pluto or Haumea).
  The extension of such studies to fainter targets is possible thanks
  to two techniques:
  a laser beam can be projected into the
  atmosphere %, which excites sodium atoms at about 90 km altitude,
  to create an artificial star of magnitude $m_R \sim 13.4$,
  called a Laser Guide Star (LGS).
  However, because the laser beam 
  is deflected on its way up by the atmospheric turbulence,
  the LGS position moves on the plane of the sky in a random pattern
  (corresponding to low orders of the turbulence, called tip-tilt).
  Hence, a natural close-by star must be monitored to correct the
  wavefront from the motion of the LGS.
  Because the requirement on these reference stars (called Tip-Tilt Star: TTS)
  are less strict (angular distance and brightness) than for NGS, several TNOs
  have already been observed this way
  \citep[\eg,][]{2006-ApJ-639-Brown, 2011-AA-528-Dumas}.

\begin{figure}
  \centering
  \includegraphics[width=.22\textwidth]{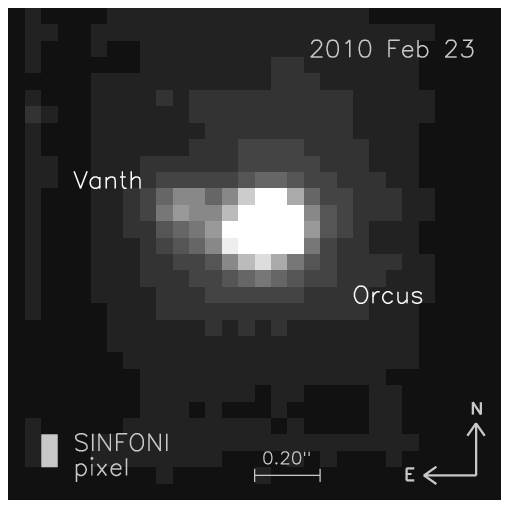}
  \includegraphics[width=.22\textwidth]{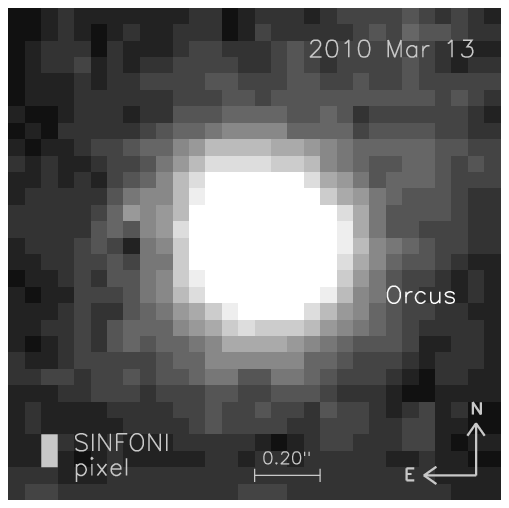}
  \caption{Two images of the Orcus-Vanth system, obtained by
    summing of all individual observations and stacking the resulting
    cube along wavelength.
    \textsl{Left:} 
    2010 February 23 UT observations using the appulse NGS AO correction,
    representing a total integration time of 4050s.
    Both Orcus and Vanth are easily separable in the image.
    \textsl{Right:}
    2010 March 13 UT observations using the LGS AO correction, without
    TTS reference (see text),
    corresponding to a total integration time of 5400s.
    The angular resolution provided in that mode forbids the detection
    of Vanth, which flux, spread over many pixels,
    is hidden within the background noise.
  } 
  \label{fig: image}
\end{figure}

  Another solution consists of computing close encounters (separation
  smaller than about 30\arcsec) on the plane
  of the sky between the object of interest and a star suitable as a 
  NGS \citep[\eg,][]{2001-DPS-33-Berthier}.
  These events are called \textsl{appulses}.
  On 2010 February 23 UT, Orcus had a particularly
  favorable appulse with 
  the star UCAC2 29643541 ($m_R = 11.6$) 
  at an angular separation of only 4\arcsec.
  We thus observed it in Service Mode
  (program ID: 
  \href{http://archive.eso.org/wdb/wdb/eso/eso_archive_main/query?prog_id=284.C-5044\%28A\%29&max_rows_returned=1000}
       {284.C-5044}) at the
  European Southern Observatory (ESO)
  Very Large Telescope (VLT) with the 
  near-infrared integral-field spectrograph 
  SINFONI
  \citep{2003-SPIE-1548-Eisenhauer, 
    2004-Msngr-117-Bonnet}.
  Observations were realized simultaneously in the atmospheric 
  H and K bands (1.45--2.45 $\mu$m) using the 
  H+K grating of SINFONI, providing a spectral resolving power $R$ of
  about 1500.
  We used a plate scale of 50$\times$100 mas/pixel, associated with a
  3\arcsec\,$\times$\,3\arcsec~field of view.
  We alternated observations of Orcus and nearby sky in a jitter pattern to
  allow optimal sky 
  correction, being cautious to avoid the NGS (4\arcsec).\\
  \indent Unfortunately, the AO module of SINFONI has not been designed to
  offer differential tracking
  (\ie, NGS fixed on the plane of the sky, field of
  view following a target with non-sidereal motion).
  We thus had to set the duration of integrations as a compromise between the
  slew of Orcus on the detector plane and the count level reached on
  Vanth ($m_V \sim 21.6$). 
  We used individual exposures of 150s to
  theoretically\,\footnote{computation made using ESO
    \href{http://www.eso.org/observing/etc/bin/gen/form?INS.MODE=swspectr+INS.NAME=sinfoni}{Exposure
    Time Calculator}}
  achieve an average signal-to-noise ratio of 1
  on Vanth over H band.
  In return, during a single exposure, Orcus moved by
  -0.109\arcsec~in right ascension and
  0.039\arcsec~in declination,
  distorting its apparent shape, thus elongated along the SE-NW
  direction as visible in Fig.~\ref{fig: image}. \\
  \indent Atmospheric conditions at the time of the observations
  were very good, with an average seeing of 0.8\arcsec~and
  a coherence time ranging from 7 to 20 ms.
  Orcus was close to zenith during the observations with an
  airmass ranging from 1.05 to 1.4.
  This allowed the AO system to provide optimal correction,
  resulting in a spatial resolution close to
  the diffraction limit of the telescope
  (Orcus FWHM was of $85\times100$ mas in K-band).\\
  \indent We also report here on some test observations of Orcus
  performed on 2010 March 13 UT at the ESO VLT
%  (prog. ID: 384.C-0877)
   (prog. ID: \href{http://archive.eso.org/wdb/wdb/eso/eso_archive_main/query?prog_id=384.C-0877\%28A\%29&max_rows_returned=1000}
           {384.C-0877})
  in the so-called ``\textsl{Seeing Enhancer}'' mode.
  This mode consists of closing the AO loop on a LGS, but without
  providing any TTS,
  Orcus itself being too faint ($m_V \sim 19.7$) to be used as TTS
  \citep[as opposed to targets like Haumea, see][for
    instance]{2011-AA-528-Dumas}. 
  Hence, only the higher orders of the atmospheric turbulence are
  corrected (\ie, there is no tip-tilt correction).
  The advantage in that mode was to perform differential
  tracking and to take longer exposures (600s).
  The instrument settings and observing strategy were otherwise
  similar to those for February observations.\\
  \indent Atmospheric conditions were worse during March
  observations, with an average seeing
  of 0.9\arcsec, and coherence time of about 3 ms.
  However, the quality of the correction provided by the AO in that
  mode is intrinsically lower than in 
  the case of the appulse: Orcus FWHM was of 0.38\arcsec~(still
  representing an improvement by a factor of $\approx$2 with respect to
  seeing-limited observations). 
  Despite the shape of Orcus being elongated by its
  apparent displacement in February, the quality of the data was superior (with a shorter
  exposure time) to the March data where the spread of its light is directly
  related to the lower AO correction achieved.
  This highlights the advantage of searching for favorable appulses
  for faint moving targets to use bright NGS as reference for the
  adaptive optics correction.
%
%
%
%
%%%%%%%%%%%%%%%%%%%%%%%%%%%%%%%%%%%%%%%%%%%%%%%%%%%%%%%%%%%%%%%%%
%%%%%%%%%%%%%%%%%%%%%%%%%%%%%%%%%%%%%%%%%%%%%%%%%%%%%%%%%%%%%%%%%
%%%%%%%%%%%%%%%%%%%%%%%%%%%%%%%%%%%%%%%%%%%%%%%%%%%%%%%%%%%%%%%%%
%%%%%%%%%%%%%%%%%%%%%%%%%%%%%%%%%%%%%%%%%%%%%%%%%%%%%%%%%%%%%%%%%
%%%%%%%%%%%%%%%%%%%%%%%%%%%%%%%%%%%%%%%%%%%%%%%%%%%%%%%%%%%%%%%%%
%%%%%%%%%%%%%%%%%%%%%%%%%%%%%%%%%%%%%%%%%%%%%%%%%%%%%%%%%%%%%%%%%

%------------------------------------------------%
%------------------------------------------------%
%---- TAG ---------------------------------------%
%------------------------------------------------%
%------------------------------------------------%
\section{Data reduction and analysis\label{sec: reduc}}
  \indent We used the SINFONI pipeline
  \citep{2007-arXiv-Modigliani}
  version 2.0.5 to perform the basic data reduction:
  bad pixel removal, flat fielding correction, 
  subtraction of the sky background from the jittered observations,
  and wavelength calibration with 
  Xenon-Argon-Krypton lamps
  \citep[see][for a complete description of the procedure on other faint
    TNOs]{2009-Icarus-201-Guilbert}.
  Default parameters were used except for in the ``\textsl{jitter}''
  recipe where 
  we set the parameters \textit{scales.sky} to true and \textit{density}
  to 3 for optimal sky-background correction.   
  This provided us with 27 and 9 individual cubes (two spatial plus
  one spectral dimensions) of Orcus/Vanth for February and March
  observations, respectively. \\
  \indent We then computed the average centroid position of Orcus for
  each individual observation by stacking the cubes along wavelength.
  We used this
  information to shift and add all the cubes into a single one for each date,
  corresponding to equivalent exposure times of 4050s and 5400s.
  We then re-aligned all the wavelength slices of the cube because
  the centroid position of Orcus was not constant with wavelength but rather
  presented a slow drift due to the differential atmospheric refraction
  as described in \citet{2010-Icarus-205-Carry-b}.\\
  \indent We then extracted the respective spectra
  of Orcus and Vanth by adjusting
  \citep[using MPFIT least-square method from][]{2009-ASPC-411-Markwardt},
  for each wavelength, a model $\mathcal{I}$
  composed by a linear background and two
  Moffat functions describing both components:
\begin{equation}
  \mathcal{I}(x,y) = \mathcal{F}_o(x,y) + \mathcal{F}_v(x,y) + ax + by + c\label{eq: fit}\\
\end{equation}
\noindent where $\mathcal{F}_i$ are the two Moffat functions,
representing Orcus ($\mathcal{F}_o$) and Vanth ($\mathcal{F}_v$), defined by:
\begin{equation}
  \mathcal{F}_i (x,y)= f_i .
     \left[
       \begin{array}{ccl}
          1 &+& \left( (x - x_i^c)\frac{\cos \theta}{\sigma_x} 
                   -   (y - y_i^c)\frac{\sin \theta}{\sigma_y} \right)^2 \\
            &+& \left( (x - x_i^c)\frac{\sin \theta}{\sigma_x} 
                   +   (y - y_i^c)\frac{\cos \theta}{\sigma_y} \right)^2 
       \end{array}
     \right]^{-\alpha}
  \nonumber
\end{equation}
  \noindent where
  $x$, $y$ are the frame spatial dimensions, 
  $f_i$ is the peak level of each Moffat function centered on 
  the coordinates ($x_i^c$, $y_i^c$).
  $\sigma_x$, $\sigma_y$ are the Half-Width at Half Max
  (HWHM) along two perpendicular directions, making an angle $\theta$
  with the detector $x$ direction, and
  $\alpha$ is the power law index of the Moffat functions. 
  The final spectrum was cleaned for bad points using a 3\,$\sigma$
  median smoothing procedure. \\
  \indent The advantage of this method is to provide the spectra of
  both components as well as their relative astrometry. We discuss
  both points in following sections.

%
%______________________________________________________________

%------------------------------------------------%
%------------------------------------------------%
%---- TAG ---------------------------------------%
%------------------------------------------------%
%------------------------------------------------%
\section{Spectral Analysis\label{sec: spec}}
  \subsection{The surface composition of Orcus}
\begin{figure}
  \centering
  \includegraphics[width=.5\textwidth]{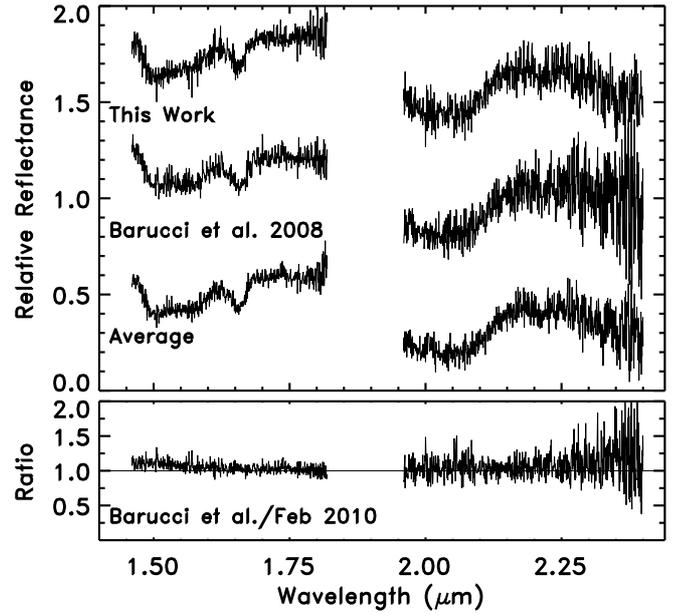}
  \caption{%
    \textsl{Upper panel}: 
    Spectra of Orcus in the H and K bands
    from this work (top) and \citet{2008-AA-479-Barucci} (middle).
    The bottom spectrum is an average of the two. The spectra are
    normalized to 1.0 at 1.75\,$\mu$m and are shifted by +0.65, 0 and
    -0.4. The spectrum taken in March 2010 was significantly
    noisier than the two shown here so we do not plot it nor do we use
    it in our analysis. 
    \textsl{Lower panel}: 
    The ratio of the \citeauthor{2008-AA-479-Barucci}
    data and data from this work
    which shows little difference between the spectra apart from a
    small flux difference shortward of 1.65\,$\mu$m.
    \label{fig: specglobal}
  }
\end{figure}

    \indent Figure~\ref{fig: specglobal} compares our new spectrum of
    Orcus to that of \citet{2008-AA-479-Barucci}.
    The overall spectral shape reveals the presence of water ice,
    dominated by the crystalline form as already addressed in previous work
    \citep{2004-AA-422-Fornasier, 
      2005-AA-437-deBergh, 
      2005-ApJ-627-Trujillo, 
      2008-AA-479-Barucci,
      2009-Icarus-201-Guilbert,
      2010-AA-520-Delsanti,
      2010-AA-521-DeMeo}.\\
    \indent The ratio of these two spectra,
    shown in the bottom part of Fig.~\ref{fig: specglobal}, reveals their similarity,
    although we note a difference in the overall flux level ($\sim$10\%)
    shortward of $\sim$1.65\,\micron.
    This difference
    does not appear to be related to variation of H$_2$O  (amount or grain size)
    because it is present shortward of 1.5\,\micron. 
    Potential explanations include instrumental effects or differences
    of the standard stars. \\
    \indent We confirm the presence of a feature near
    2.2\,\micron~with a band center located at  
    2.209\,$\pm$\,0.002\,\micron~and a band depth of 9\,$\pm$\,2\% 
    that previous works have attributed to
    CH$_4$, NH$_3$, or NH$_4^+$.
    We combined the \citeauthor{2008-AA-479-Barucci} spectrum with ours to
    slightly increase the 
    overall signal-to-noise ratio. We did not include the spectra from
    \citet{2010-AA-520-Delsanti} 
    or \citet{2010-AA-521-DeMeo} because the quality of those data
    were significantly lower.  
    This average spectrum is used for 
    all of the analysis reported in this section. \\
%
%
%%%%%%%%%%%%%%%%%%%%%%%%%%%%%%%%%%%%%%%%%%%%%%%%%%%%%%%%%%%%%%%%%
%%%%%%%%%%%%%%%%%%%%%%%%%%%%%%%%%%%%%%%%%%%%%%%%%%%%%%%%%%%%%%%%%
%%%%%%%%%%%%%%%%%%%%%%%%%%%%%%%%%%%%%%%%%%%%%%%%%%%%%%%%%%%%%%%%%
%%%%%%%%%%%%%%%%%%%%%%%%%%%%%%%%%%%%%%%%%%%%%%%%%%%%%%%%%%%%%%%%%
%%%%%%%%%%%%%%%%%%%%%%%%%%%%%%%%%%%%%%%%%%%%%%%%%%%%%%%%%%%%%%%%%
%%%%%%%%%%%%%%%%%%%%%%%%%%%%%%%%%%%%%%%%%%%%%%%%%%%%%%%%%%%%%%%%%
%
%
    \indent Current volatile retention models
    \citep[\eg,][]{2007-ApJ-659-Schaller,2009-Icarus-202-Levi}
    predict that CH$_4$ is not stable on Orcus' surface over its lifetime,
    however, Orcus' intermediate size among TNOs
    place it closer to the retention boundary than most other
    objects and provides us with an opportunity to test these models
    and perhaps make constraints on the assumptions therein. 
    An important step in understanding Orcus' surface 
    composition is thus a search for weak bands hidden in the
    spectrum near the detection limit. While many species could potentially exist in small
    quantities on the surface, the lack of multiple strong bands makes
    their identification difficult. Here we focus on searching for
    methane bands in the spectrum primarily because of the abundance
    of strong bands in the appropriate wavelength regime, but also
    because of the important implications a detection would have on
    our understanding of the surface conditions of these bodies and the 
    criteria for volatile retention. \\
%
%
%
%
%%%%%%%%%%%%%%%%%%%%%%%%%%%%%%%%%%%%%%%%%%%%%%%%%%%%%%%%%%%%%%%%%
%%%%%%%%%%%%%%%%%%%%%%%%%%%%%%%%%%%%%%%%%%%%%%%%%%%%%%%%%%%%%%%%%
%%%%%%%%%%%%%%%%%%%%%%%%%%%%%%%%%%%%%%%%%%%%%%%%%%%%%%%%%%%%%%%%%
\begin{figure}
  \centering
  \includegraphics[width=.5\textwidth]{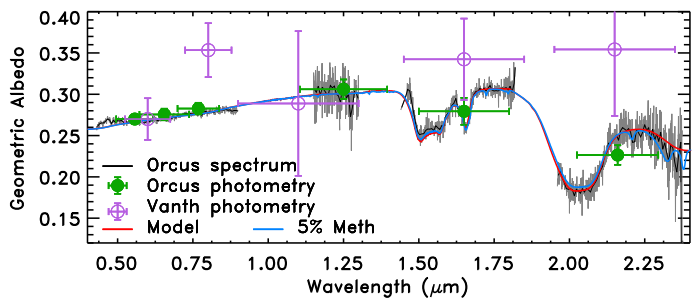}
  \caption{%
    Plotted here is the average spectrum of Orcus from our work and
    \citeauthor{2008-AA-479-Barucci}
    together with the visible data from
    \citet{2009-AA-493-DeMeo}, scaled to the visible albedo estimated by
    \citet{2010-AA-518-Lim}.
    Overplotted are the basic model in red
    (\#1 in Table~\ref{tab: model})
    and model with additional 5\% methane in blue (\#2, \textsl{ibid}).
    Also shown are the visible
    and near-infrared colors of Vanth in purple.
    V, I and J band measurements are from
    \citet{2010-AJ-139-Brown}. {The colors of Vanth are normalized to Orcus' spectrum at 0.6 microns.}
  } 
  \label{fig: vanth}
\end{figure}
%%%%%%%%%%%%%%%%%%%%%%%%%%%%%%%%%%%%%%%%%%%%%%%%%%%%%%%%%%%%%%%%%
%%%%%%%%%%%%%%%%%%%%%%%%%%%%%%%%%%%%%%%%%%%%%%%%%%%%%%%%%%%%%%%%%
%%%%%%%%%%%%%%%%%%%%%%%%%%%%%%%%%%%%%%%%%%%%%%%%%%%%%%%%%%%%%%%%%
%
%
%
    \indent First, to remove the dominant signature of the crystalline
    water ice from the spectrum we 
    model the composition of Orcus using a code
    based on radiative transfer theory
    \citep{hapke-theory} using optical constants of laboratory materials
    for inputs.
    {We use optical constants of H$_2$O ice in both
      crystalline  and amorphous form (at 40\,K and 38\,K from
      \citealt{1998-JGR-103-Grundy} and 
     \citealt{1998-ASSL-227-Schmitt}, respectively) and Titan and 
     Triton Tholin
     \citep{1984-Icarus-60-Khare, 1993-Icarus-103-Khare}.
     The temperature of the H$_2$O optical
     constants are appropriate because the blackbody temperature at
     39\,AU is about 43\,K and Pluto's surface temperature (with
     a similar semi-major axis) is measured  
     to be 40\,$\pm$\,2\,K \citep{1994-Icarus-112-Tryka}.
     Triton and Titan
     Tholins are used as representative material  
     that aid in fitting the spectrum, because optical constants are
     available for these materials.  
     However they could be replaced with different materials, such as
     other organics that have 
     similar spectral properties.
 }
    The models we present here
    (see Table~\ref{tab: model} and Fig.~\ref{fig: vanth})
    {are based on the recent analysis by
    \citet{2010-AA-521-DeMeo}} but differ slightly owing to
    the recent reevaluation of the albedo of Orcus 
    from 0.20\,$\pm$\,0.03 \citep{2008-SSBN-3-Stansberry}
    to 0.27\,$\pm$\,0.06 by \citet{2010-AA-518-Lim}.
    We reduced the amount of amorphous water ice to about 10\%, 
    increasing the crystalline H$2$O by the same amount to
    better fit the 1.65$\mu$ band and adjusted
    the cosine asymmetry factor \citep{hapke-theory} 
    to properly fit the data's higher albedo.
    The spectrum was then divided by the model 1, without methane 
    (Table~\ref{tab: model}).
    We created a program in IDL designed to fit
    Gaussians to potential features in designated wavelength
    regions. The program was set to search in the regions near 1.67,
    1.72, and 2.2\,\micron, where methane absorbs strongly
    \citep{1997-Icarus-127-Quirico}. We did not search for the band
    near 1.80\,\micron~because of poor telluric correction in this
    wavelength range, 
    nor the bands near 2.32\,\micron~and 2.43\,\micron~due to
    decreasing signal-to-noise ratio at 
    wavelengths longer than $\approx$2.3\,\micron~from low detector
    sensitivity.
    A least-square
    minimization \citep{2009-ASPC-411-Markwardt}
    was used to find the best-fit center, width, and
    depth of the bands. \\
%
%%%%%%%%%%%%%%%%%%%%%%%%%%%%%%%%%%%%%%%%%%%%%%%%%%%%%%%%%%%%%%%%%
%%%%%%%%%%%%%%%%%%%%%%%%%%%%%%%%%%%%%%%%%%%%%%%%%%%%%%%%%%%%%%%%%
%%%%%%%%%%%%%%%%%%%%%%%%%%%%%%%%%%%%%%%%%%%%%%%%%%%%%%%%%%%%%%%%%
\begin{table}
\caption{Material present in the models of surface composition.} 
\label{tab: model}
\centering
\begin{tabular}{clcc}
\hline\hline
&\multicolumn{1}{c}{Material} & Amount (\%) & Grain size (\micron) \\
\hline
 Model 1 & Crystalline H$_2$O & 60\textcolor{white}{.0}   & \textcolor{white}{1}18 \\
         & Amorphous H$_2$O   & 10.5 & \textcolor{white}{1}18 \\
         & Titan tholins      & \textcolor{white}{1}2\textcolor{white}{.0}   & \textcolor{white}{1}10 \\
         & Triton tholins     & \textcolor{white}{1}6\textcolor{white}{.0}   & \textcolor{white}{1}10 \\
         & Blue compound      & 22.5 & \textcolor{white}{1}10 \\
     \noalign{\smallskip}
 Model 2 & Crystalline H$_2$O & 60\textcolor{white}{.0}   & \textcolor{white}{1}18 \\
         & Amorphous H$_2$O   & \textcolor{white}{1}5.5 & \textcolor{white}{1}18 \\
         & Titan tholins      & \textcolor{white}{1}2\textcolor{white}{.0}   & \textcolor{white}{1}10 \\
         & Triton tholins     & \textcolor{white}{1}6\textcolor{white}{.0}   & \textcolor{white}{1}10 \\
         & Blue compound      & 22.5 & \textcolor{white}{1}10 \\
         & Methane            & \textcolor{white}{1}5\textcolor{white}{.0}   & 100\\
\hline
\end{tabular}

\end{table}
%%%%%%%%%%%%%%%%%%%%%%%%%%%%%%%%%%%%%%%%%%%%%%%%%%%%%%%%%%%%%%%%%
%%%%%%%%%%%%%%%%%%%%%%%%%%%%%%%%%%%%%%%%%%%%%%%%%%%%%%%%%%%%%%%%%
%%%%%%%%%%%%%%%%%%%%%%%%%%%%%%%%%%%%%%%%%%%%%%%%%%%%%%%%%%%%%%%%%

\begin{figure}
  \centering
  \includegraphics[width=.5\textwidth]{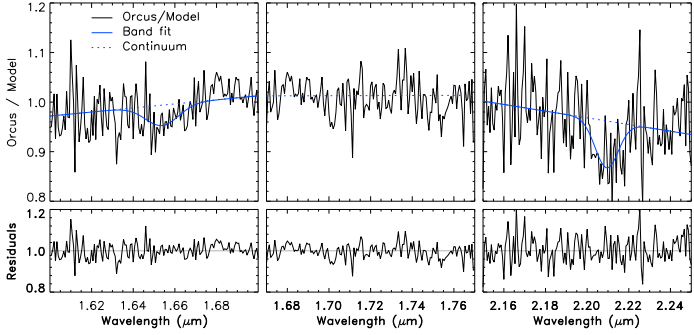}
  \caption{%
    Gaussian fits to the data to search for weak bands.
    Two bands of approximately 4\% and 9\% are detected 
    around 1.65 and 2.2\,\micron~respectively (see
    Table~\ref{tab: bands}).
    No band with depth weaker than about 2\% is detectable around
    1.72\,\micron. 
  } 
  \label{fig: bands}
\end{figure}

    \indent 
    The results of the Gaussian fits are listed in
    Table~\ref{tab: bands}, and plots of the fits are shown in Fig.~\ref{fig: bands}. 
    We find a Gaussian fit near the 1.67\,\micron~feature, however,
    the center is at 1.654\,\micron~indicating it is residual
    crystalline H$_2$O not removed by the 
    division of the data by the model.  
    We do not find
    a band at 1.67 \micron~nor at 1.72\,\micron. 
    The depth of the 2.209\,\micron~Gaussian fit is 9.5\,$\pm$\,2.3\%.  
    While we do not fit a Gaussian to the 2.32\,\micron~feature we do
    note that the reflectance of the data is lower in this region
    with respect to the model that excludes
    CH$_4$ (\#1) suggesting there is a compound  
    absorbing in this region, though not necessarily CH$_4$. \\
    \indent     
    The question that remains is
    ``Can the band at 2.2\,\micron~be due to methane even 
    though we do not detect other bands?'' The relative band depths of
    the 1.67 and 1.72\,\micron~features compared to the
    2.2\,\micron~feature are 95\% and 88\%,  
    respectively, assuming a grain  
    size of 200\,\micron~(the relative depths do not change
    significantly with variations in grain size of
    $\pm$100\,\micron). 
    We should clearly detect all three bands if the
    2.2\,\micron~feature were due to methane only
    (as visible in Fig~\ref{fig: vanth} by comparing the spectrum
    of Orcus with model \#2, including 5\% of methane to account
    for the 2.2\,\micron~feature), 
    thus we place a limit of a maximum of about 2\% of methane on 
    the surface. Our constraint of 2\% is lower than the
    constraint set in \citet{2010-AA-521-DeMeo} partly 
    because of the better quality of the data, but also because bands
    from 5\% methane model at the higher albedo  
    are weaker than for the lower albedo model.
    The majority (or all)
    of the 2.2\,$\mu$m absorption must 
    therefore be due to either hydrated ammonia, ammonium, as suggested  
    in previous work, or another yet unknown compound that absorbs in
    this region. Since H$_2$O:NH$_3$ and NH${_4^+}$ do not  
    have any other distinguishing features in the wavelength range of
    our data we cannot  
    confirm or exclude their presence.\\
    \indent       
    We also search for bands at 2.274 and 2.314\,\micron, the positions 
    of the strongest bands of ethane
    \citep{1997-Icarus-127-Quirico}. Ethane is a by-product of methane 
    and its potential presence on Orcus was suggested by models to
    data from \citet{2010-AA-520-Delsanti}. 
    We do not find any bands at the locations of ethane absorption
    within the limits of the  
    quality of our data (which is the best to date). Either there is
    no ethane (less than about 5\%, depending on grain size, within
    the limits of detection  
    with our data) on Orcus' surface or, less likely,
    there is a concentration of
    ethane on the part of the surface observed by  
    \citeauthor{2010-AA-520-Delsanti} that we did not observe. We also
    check for the possible presence of other 
    volatile compounds: CO$_2$ (three narrow features near 2\,\micron),
    CO (1.578\,\micron, we do not search at 2.352\,$\mu$m  
    because of the poor quality of the data), N$_2$ (2.15\,\micron), and
    methanol (2.27\,\micron). We do not detect any of these  
    features in our data.
%    within the limits of the signal-to-noise ratio.

%
%
%%%%%%%%%%%%%%%%%%%%%%%%%%%%%%%%%%%%%%%%%%%%%%%%%%%%%%%%%%%%%%%%%
%%%%%%%%%%%%%%%%%%%%%%%%%%%%%%%%%%%%%%%%%%%%%%%%%%%%%%%%%%%%%%%%%
%%%%%%%%%%%%%%%%%%%%%%%%%%%%%%%%%%%%%%%%%%%%%%%%%%%%%%%%%%%%%%%%%
\begin{table}
\caption{Parameters for weak bands.} 
\label{tab: bands}
\centering
\begin{tabular}{lcc@{ $\pm$ }cc@{ $\pm$ }cc@{ $\pm$ }c}
\hline\hline
\multicolumn{1}{c}{Species} & $\lambda_e$ &
\multicolumn{2}{c}{$\lambda_m$} &
\multicolumn{2}{c}{$\Delta \lambda$} &
\multicolumn{2}{c}{Depth} \\
& (\micron) &
\multicolumn{2}{c}{(\micron)} &
\multicolumn{2}{c}{(nm)} & 
\multicolumn{2}{c}{(\%)} \\
\hline
CH$_4$     & 1.670 & 1.654 & 0.004 &
                         8 & 5 &
                       4.1 & 1.9 \\ 
CH$_4$     & 1.724 & \multicolumn{2}{c}{--} &
                     \multicolumn{2}{c}{--} &
                     \multicolumn{2}{c}{$ < 2$} \\
CH$_4$     & 2.208 & 2.209 & 0.002 &
                         6 & 2 &
                       9.5 & 2.3 \\
C$_2$H$_6$ & 2.274 & \multicolumn{2}{c}{--} &
                     \multicolumn{2}{c}{--} &
                     \multicolumn{2}{c}{$ < 2$} \\
C$_2$H$_6$ & 2.314 & \multicolumn{2}{c}{--} &
                     \multicolumn{2}{c}{--} &
                     \multicolumn{2}{c}{$ < 2$} \\
\hline
\end{tabular}
\tablefoot{
  Band centers ($\lambda_e$) of the strongest CH$_4$ bands from
  \citet{1997-Icarus-127-Quirico}. We list the band center ($\lambda_m$),
  width ($\Delta \lambda$), and depth, from the Gaussian fit to our
  data at these wavelengths if detected.} 
\end{table}
%%%%%%%%%%%%%%%%%%%%%%%%%%%%%%%%%%%%%%%%%%%%%%%%%%%%%%%%%%%%%%%%%
%%%%%%%%%%%%%%%%%%%%%%%%%%%%%%%%%%%%%%%%%%%%%%%%%%%%%%%%%%%%%%%%%
%%%%%%%%%%%%%%%%%%%%%%%%%%%%%%%%%%%%%%%%%%%%%%%%%%%%%%%%%%%%%%%%%
%
%%%%%%%%%%%%%%%%%%%%%%%%%%%%%%%%%%%%%%%%%%%%%%%%%%%%%%%%%%%%%%%%%
%%%%%%%%%%%%%%%%%%%%%%%%%%%%%%%%%%%%%%%%%%%%%%%%%%%%%%%%%%%%%%%%%
%%%%%%%%%%%%%%%%%%%%%%%%%%%%%%%%%%%%%%%%%%%%%%%%%%%%%%%%%%%%%%%%%
%%%%%%%%%%%%%%%%%%%%%%%%%%%%%%%%%%%%%%%%%%%%%%%%%%%%%%%%%%%%%%%%%
%%%%%%%%%%%%%%%%%%%%%%%%%%%%%%%%%%%%%%%%%%%%%%%%%%%%%%%%%%%%%%%%%
%%%%%%%%%%%%%%%%%%%%%%%%%%%%%%%%%%%%%%%%%%%%%%%%%%%%%%%%%%%%%%%%%
%

  \subsection{The colors of Vanth}
    \indent The spectrum of Vanth we extracted
    had a very low signal-to-noise ratio,
    therefore we drastically reduced its spectral resolution
    and we only report in
    Table~\ref{tab: colors} its H- and K-band magnitude difference to Orcus.
    Fig.~\ref{fig: vanth} compares the visible and near-infrared
    colors of Vanth to those of Orcus, scaled to the visible
    albedo of Orcus from \citet{2010-AA-518-Lim}. 
    Although the errors are large, the colors indicated that Vanth 
    is both slightly redder in the visible wavelength regime as reported by
    \citet{2010-AJ-139-Brown} and the near-infrared indicated by our data.
    While the colors of Vanth do not suggest a strong presence of water ice
    as seen on Orcus, the possibility cannot be ruled out within the 
    uncertainty of the data.
    As already addressed by \citeauthor{2010-AJ-139-Brown}, it
    is not possible to draw any conclusions on the origin of this
    color difference to distinguish 
    between a capture or collision formation scenario.
    Even if Vanth is confirmed to have 
    colors inconsistent with water ice it may not exclude a collision
    formation scenario. 
    
    %http://www.lpi.usra.edu/publications/absearch/?meeting=335&keywords_all=Cook&submit.search=Search
    %http://adsabs.harvard.edu/abs/2011arXiv1103.2175B

%
%
%
%%%%%%%%%%%%%%%%%%%%%%%%%%%%%%%%%%%%%%%%%%%%%%%%%%%%%%%%%%%%%%%%%
%%%%%%%%%%%%%%%%%%%%%%%%%%%%%%%%%%%%%%%%%%%%%%%%%%%%%%%%%%%%%%%%%
%%%%%%%%%%%%%%%%%%%%%%%%%%%%%%%%%%%%%%%%%%%%%%%%%%%%%%%%%%%%%%%%%
\begin{table}
\caption{Relative magnitude difference of Orcus and Vanth.}
\label{tab: colors}
\centering
\begin{tabular}{ccccc}
\hline\hline
Filter & Orcus$^{1}$ & $\Delta$mag & Flux ratio & Refs.\\
\hline
V  & 19.36 $\pm$ 0.05 & 2.61 $\pm$ 0.10 &  9.0 $\pm$ 0.9 & 1\\
I  & 18.63 $\pm$ 0.05 & 2.31 $\pm$ 0.10 &  8.6 $\pm$ 2.9 & 1\\
J  & 20.64 $\pm$ 0.04 & 2.66 $\pm$ 0.34 & 11.5 $\pm$ 0.5 & 1\\
H  & 21.65 $\pm$ 0.03 & 2.34 $\pm$ 0.12 & 11.6 $\pm$ 1.3 & 2\\
Ks &        --        & 2.07 $\pm$ 0.24 & 14.9 $\pm$ 3.2 & 2\\
\hline
\end{tabular}
\tablebib{
(1)~\citet{2010-AJ-139-Brown}, (2)~This work.
}
\end{table}
%%%%%%%%%%%%%%%%%%%%%%%%%%%%%%%%%%%%%%%%%%%%%%%%%%%%%%%%%%%%%%%%%
%%%%%%%%%%%%%%%%%%%%%%%%%%%%%%%%%%%%%%%%%%%%%%%%%%%%%%%%%%%%%%%%%
%%%%%%%%%%%%%%%%%%%%%%%%%%%%%%%%%%%%%%%%%%%%%%%%%%%%%%%%%%%%%%%%%

%------------------------------------------------%
%------------------------------------------------%
%---- TAG ---------------------------------------%
%------------------------------------------------%
%------------------------------------------------%
\section{Orbital characterization\label{sec: astro}}
\begin{figure}
  \centering
  \includegraphics[width=.5\textwidth]{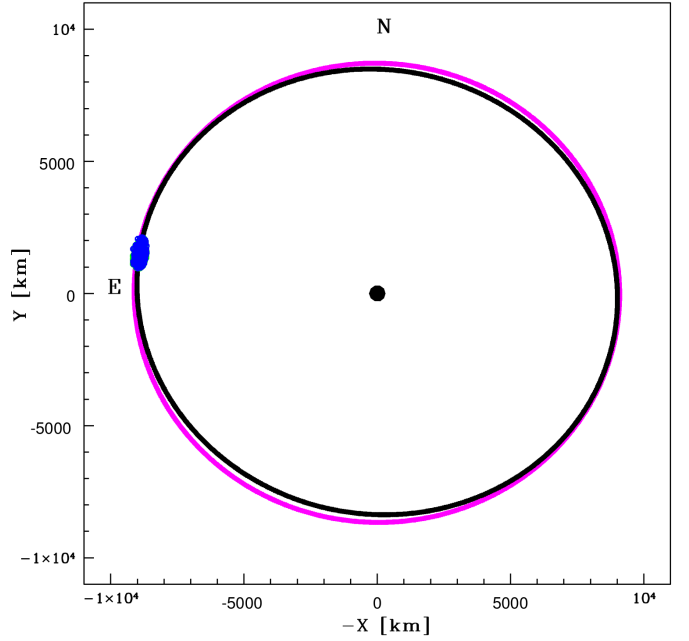}
  \caption{
    Predicted position of Vanth relative to Orcus on 2010 Feb. 23.237,
    from the extrapolation of the initial 2004--2007 data by
    \citet{2010-AJ-139-Brown}.
    The two curves reprensent two possible solutions with
    symmetric pole solutions. The dots in the Eastern part of the frame
    correspond to the predicted positions at the time of the SINFONI
    observation for all retained orbital solutions given in
    Fig.~\ref{fig: aeip}
  } 
  \label{fig: pred}
\end{figure}

  \indent The relative orbit of the Orcus/Vanth system has been determined
  previously by \citet{2010-AJ-139-Brown} from observations conducted
  on the Hubble Space Telescope (HST), mainly with
  the High Resolution Channel
  of the Advanced Camera for Surveys (ACS),
  acquired over the period 2005--2007.   
  Three years after, 
  the uncertainty in the orbital period
  leads to a position uncertainty of Vanth along its orbit of about 
  15\degr, corresponding to 2,000\,km (Fig.~\ref{fig: pred}).
  With one additional point for the relative position of the components
  after 3 years, corresponding to some 2,000 revolutions of the pair, it
  is possible to refine the orbit, especially the orbital period,
  which in turn allows 
  refining
  the position prediction
  for future observations, in particular stellar occultations.\\
  \indent We present in this section the relative astrometry
  measurements, the subsequent orbit computation,
  and the consequences on predictions of stellar occultations.

%%%%%%%%%%%%%%%%%%%%%%%%%%%%%%%%%%%%%%%%%%%%%%%%%%%%%%%%%%%%%%%%%
%%%%%%%%%%%%%%%%%%%%%%%%%%%%%%%%%%%%%%%%%%%%%%%%%%%%%%%%%%%%%%%%%
%%%%%%%%%%%%%%%%%%%%%%%%%%%%%%%%%%%%%%%%%%%%%%%%%%%%%%%%%%%%%%%%%
%%%%%%%%%%%%%%%%%%%%%%%%%%%%%%%%%%%%%%%%%%%%%%%%%%%%%%%%%%%%%%%%%
%%%%%%%%%%%%%%%%%%%%%%%%%%%%%%%%%%%%%%%%%%%%%%%%%%%%%%%%%%%%%%%%%
%%%%%%%%%%%%%%%%%%%%%%%%%%%%%%%%%%%%%%%%%%%%%%%%%%%%%%%%%%%%%%%%%

  \subsection{Astrometry measurements}
    \indent To extract the relative astrometry of Orcus and Vanth,
    we first stacked the spectrocubes along the wavelength
    dimension to obtain two frames of higher SNR, one at each epoch.
    We then used the image model $\mathcal{I}$ presented in
    Sect.~\ref{sec: reduc}, the astrometry being given by
    the relative position of the Moffat functions centers.\\
    \indent As visible in Fig.~\ref{fig: image}, Vanth is easily
    separable from Orcus on the image obtained in February, and we
    estimate the accuracy of its position to be about 25 mas,
    corresponding to a fourth of a pixel on the SINFONI detector.
    However, our observing date, which was imposed by the stellar appulse,
    was not optimal for taking full advantage of the parallactic effect
    and removing the ambiguity in the relative orbit inclination
    (see the two curves in Fig.~\ref{fig: pred}). On the other hand,
    the moderate spatial resolution (FWHM 
    of 380 mas) achieved in March did not allowed the localization of
    Vanth, which flux was spread over a large area and diluted in the
    background noise (Fig.~\ref{fig: image}). \\
    \indent We also use the astrometric measurements given in
    \citet{2010-AJ-139-Brown}, although, after analyzing the public HST data,
    it is clear that the dates given in
    \citet{2010-AJ-139-Brown} are wrong by a constant offset of half a 
    day (confirmed by D. Ragozzine and M. Brown, personal communication).
    All the astrometric measurements are correctly reproduced in
    Table~\ref{tab: astro}, along with the observation circumstances: 
    heliocentric distance ($\Delta$) and range to observer ($r$),
    right ascencion ($\alpha$) and declination ($\delta$).
%

%%%%%%%%%%%%%%%%%%%%%%%%%%%%%%%%%%%%%%%%%%%%%%%%%%%%%%%%%%%%%%%%%
%%%%%%%%%%%%%%%%%%%%%%%%%%%%%%%%%%%%%%%%%%%%%%%%%%%%%%%%%%%%%%%%%
%%%%%%%%%%%%%%%%%%%%%%%%%%%%%%%%%%%%%%%%%%%%%%%%%%%%%%%%%%%%%%%%%
\begin{table*}
\caption{Astrometric data for the relative position of
    Orcus/Vanth used to reconstruct their mutual orbit.}
\label{tab: astro}
\centering
\begin{tabular}{rrrrrrrc}
\hline
\hline
\noalign{\smallskip}
  \multicolumn{1}{c}{Date$^{\dagger}$} & 
  \multicolumn{1}{c}{$\Delta X^{\ddagger}$} & \multicolumn{1}{c}{$\Delta Y^{\ddagger}$} & 
  \multicolumn{1}{c}{$\Delta$} & \multicolumn{1}{c}{$r$} &
  \multicolumn{1}{c}{$\alpha$} & \multicolumn{1}{c}{$\delta$} &
  Instr.\\
  \multicolumn{1}{c}{(JD)} &
  \multicolumn{1}{c}{(\arcsec)} & \multicolumn{1}{c}{(\arcsec)}&
  \multicolumn{1}{c}{(AU)} & \multicolumn{1}{c}{(AU)} &
  \multicolumn{1}{c}{(\degr)} & \multicolumn{1}{c}{(\degr)}\\
\noalign{\smallskip}
\hline
\noalign{\smallskip}
   2453687.66400 &   0.206 &  -0.147 & 47.811 & 47.702 & 144.039 & -4.356 & ACS \\
   2454040.36900 &   0.226 &  -0.111 & 48.075 & 47.752 & 144.813 & -4.729 & ACS \\
   2454044.36600 &  -0.258 &  -0.005 & 48.013 & 47.752 & 144.839 & -4.766 & ACS \\
   2454051.57900 &  -0.006 &  -0.243 & 47.898 & 47.753 & 144.871 & -4.828 & ACS \\
   2454056.08900 &  -0.036 &   0.240 & 47.824 & 47.754 & 144.882 & -4.865 & ACS \\
   2454066.14600 &   0.053 &   0.240 & 47.659 & 47.755 & 144.882 & -4.938 & ACS \\
   2454080.33600 &  -0.030 &  -0.244 & 47.433 & 47.757 & 144.824 & -5.019 & ACS \\
   2454416.29300 &  -0.263 &  -0.024 & 47.968 & 47.801 & 145.703 & -5.304 & NICMOS3 \\
   2454439.78000 &   0.245 &   0.078 & 47.585 & 47.804 & 145.696 & -5.473 & WFPC2 \\
\noalign{\smallskip}
\hline                                                                   
\noalign{\smallskip}
   2455250.73686 &   0.260 &   0.052 & 46.954 & 47.895 & 146.287 & -6.396 &SINFONI \\
\noalign{\smallskip}
\hline
\end{tabular}
\begin{list}{}{}
\item[$^{\dagger}$] The first
    block is reproduced from \citet{2010-AJ-139-Brown} with the corrected
    values for the dates (see text).
    The second bloc correspond to our appulse observation on 2010
    February 23.
\item[$^{\ddagger}$] Positions are positive through East and North.
\end{list}
\end{table*}
%%%%%%%%%%%%%%%%%%%%%%%%%%%%%%%%%%%%%%%%%%%%%%%%%%%%%%%%%%%%%%%%%
%%%%%%%%%%%%%%%%%%%%%%%%%%%%%%%%%%%%%%%%%%%%%%%%%%%%%%%%%%%%%%%%%
%%%%%%%%%%%%%%%%%%%%%%%%%%%%%%%%%%%%%%%%%%%%%%%%%%%%%%%%%%%%%%%%%

  \subsection{Orbit improvement}
\begin{figure}
  \centering
  \includegraphics[width=.5\textwidth]{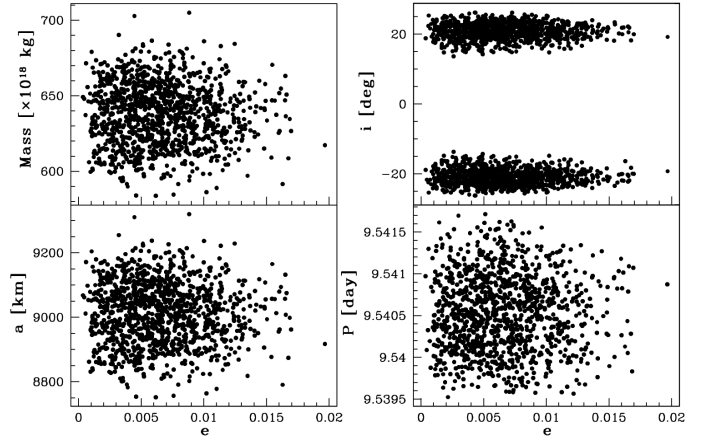}
  \caption[Orbital solutions in the elliptical elements space.]
  {
    Elliptical orbital elements:
    System mass ($\mathcal{M}$),
    semi-major axis ($a$), inclination ($i$), and
    period ($P$), plotted as function of the eccentricity ($e$).
    Each point represents a different orbital solution that fits
    the astrometry measurements in Table~\ref{tab: astro}.
    See Table~\ref{tab: orbit} for a summary of the orbital elements.
  } 
  \label{fig: aeip}
\end{figure}

    \indent With the additional astrometric data we can improve the orbital
    parameters, reset the orbital phase of Vanth,
    and, taking advantage of the time leverage, the orbital period 
%    and total mass are
    is still better constrained.
    We computed an improved Keplerian orbit
    using a statistical inversion algorithm
    \citep{2005-EMP-97-Hestroffer}, which allows to probe
    a large portion of the orbital parameters space.
    We show in Fig.~\ref{fig: aeip} all the solutions consistent with
    the observations, within the measurement uncertainties, and list
    the resulting improved orbital and physical parameters in
    Table~\ref{tab: orbit}.\\
    \indent As mentioned before,
    the position of Vanth in its orbit was not favorable on 2010 Feb
    23 (see Fig.~\ref{fig: pred}) to remove the ambiguity of the
    inclination of the orbit 
    ($\pm 21$\degr with respect to the average plane-of-the-sky).
    However, 
    the orbital pole solution that is almost
    perpendicular to the ecliptic
    (P$_{\rm A}$, corresponding to $i=+21$\degr),
    coordinates ECJ2000 (321\degr,~-2\degr),
    is statistically more probable given the observable data
    (see Fig.~\ref{fig: aeip});
    although the other solution (P$_{\rm B}$)
    inclined at about 52\degr~from the ecliptic, 
    coordinates ECJ2000 (340\degr,~+38\degr),
    provides very similar residuals and cannot be ruled out yet.\\
    \indent All the possible solutions we find have low eccentricities
    ($e$\,$\le$\,0.01). \citet{2010-AJ-139-Brown} had already highlighted
    the nearly circular nature of Vanth's orbit around Orcus, although
    we find here several solutions with larger eccentricities than the
    upper limit they reported (0.0036, 1\,$\sigma$ deviation to the
    data).
    Similarly, we find an average orbital period longer than
    \citet{2010-AJ-139-Brown}
    \citep[consistent with][however]{2011-AA-525-Ortiz},
    their shorter period being still possible, although less likely.
    These differences can originate from the circular assumption
    on Vanth orbit used by \citeauthor{2010-AJ-139-Brown}.
    In any case, since
    the eccentricity of the orbit is small,
    it suggests the tides have
    circularized the orbit, and subsequently that the satellite is also in
    synchronous rotation.
    The time for circularisation of such a system can
    be fast, of the order of  
    $\tau_{\rm c}$\,$\sim$\,$(a/R_{2})^5\,n^{-1}$\,$\sim$\,$10^6$ years
    \citep{2008-CeMDA-101-Ferraz-Mello}.
    Another reasonable consequence from such tidal evolution is that
    the orbital plane is aligned with the equator of the primary.
    In this case, we would currently see Orcus under a small
    aspect angle (roughly 10\degr~and 30\degr~for P$_{\rm A}$ and
    P$_{\rm B}$ respectively).\\
    \indent The circular orbit could also be due,
    independently of tidal effects, to
    a Kozai resonance 
    \citep[$\sqrt{1-e^2}\,\cos i$\,=\,constant,][]{1962-AJ-67-Kozai},
    in which case the
    inclination of Vanth's orbit with respect to Orcus' equator can be
    large. This means also that the direction of the spin for the
    primary is unconstrained and that its aspect angle can be much
    larger. We favor here the first solution (P$_{\rm A}$) as 
    it is slightly more probable statistically, and
    it is consistent with the low amplitude of the variations
    present in the optical lightcurve of Orcus 
    \citep[0.04 mag, see][and references
      therein]{2010-AA-522-Thirouin}.\\

%
%%%%%%%%%%%%%%%%%%%%%%%%%%%%%%%%%%%%%%%%%%%%%%%%%%%%%%%%%%%%%%%%%
%%%%%%%%%%%%%%%%%%%%%%%%%%%%%%%%%%%%%%%%%%%%%%%%%%%%%%%%%%%%%%%%%
%%%%%%%%%%%%%%%%%%%%%%%%%%%%%%%%%%%%%%%%%%%%%%%%%%%%%%%%%%%%%%%%%
%%%%%%%%%%%%%%%%%%%%%%%%%%%%%%%%%%%%%%%%%%%%%%%%%%%%%%%%%%%%%%%%%
%%%%%%%%%%%%%%%%%%%%%%%%%%%%%%%%%%%%%%%%%%%%%%%%%%%%%%%%%%%%%%%%%
%%%%%%%%%%%%%%%%%%%%%%%%%%%%%%%%%%%%%%%%%%%%%%%%%%%%%%%%%%%%%%%%%
%

    \indent The total mass $\mathcal{M}$
    is derived to high precision from Kepler's
    third law, considering this dynamical system is well
    isolated over the observations time scale, 
    supported by the two-body fit.
    Then, the recent size determination from ESA Herschel
    by \cite{2010-AA-518-Lim} allows us to
    evaluate the volume, hence density, of the components.
    Because the
    components are not resolved by the thermal infrared pixels, one can
    only derive a radiometric volume-equivalent diameter $\mathcal{D}$ for the
    system, and subsequently model-dependent values for the density. 
    This effective diameter
    can be simply related to the volume-equivalent diameter
    of each component by
    considering their projected apparent surface
    $\mathcal{D}^2$\,=\,$D^2_{\rm o}$\,+\,$D^2_{\rm v}$.
    {Assuming} that the two components are of similar
    density, one eventually gets a rough
    estimation of the bulk density:
\begin{equation}
%  \rho = \frac{3\,M}{4\,\pi\,R^3_{\rm eff}} \;
  \rho = \frac{6\,\mathcal{M}}{\pi\,\mathcal{D}^3} \;
  \frac{ (1+\mathcal{R}^2)^{3/2} }{(1+\mathcal{R}^3)} 
  \label{eq: density}
\end{equation}
    \noindent where the size ratio of the bodies
    $\mathcal{R} \equiv D_{\rm v}/D_{\rm o}= \sqrt{\frac{p_{\rm o}}{p_{\rm v}}}\,10^{-0.2\,\Delta m}$,
    essentially depend on their apparent magnitude difference
    ($\Delta m$\,$\approx$\,2.4\,$\pm$\,0.4, see Table~\ref{tab: colors}), 
    modulated by their unknown albedo ratio.
    Following \citet{2010-AJ-139-Brown}, we consider two possible
    {albedos} for Vanth: 0.27 (\ie, equals to Orcus', corresponding to
    co-formation 
    mechanisms, Section~\ref{ssec: formation}), and 0.12, typical
    for TNOs exempt of water ice (for capture mechanisms, 
    Section~\ref{ssec: formation}).
    We find mass ratios of 30 and 8, for the equal-albedo and different-albedo
    assumptions respectively, close to the estimates by
    \citet{2010-AJ-139-Brown} and
    \citet{2011-AA-525-Ortiz}.
    Based on the refined volume-equivalent diameter of the Orcus/Vanth
    system by 
    \citet{2010-AA-518-Lim}, which
    is 10\% smaller than the estimate by 
    \citet{2010-AJ-139-Brown}, 
    we find a density of about 2.3\,$\pm$\,0.8\,g/cc.
    In any case, the uncertainty on the bulk density is dominated by the
    uncertainty on the size determination, which cannot be more accurate
    than about 10\% \citep{2010-AA-518-Lim}.
    The relative precision on the bulk density cannot therefore be smaller
    than about 30\%.  
    Better knowledge of the spin properties of Orcus, and of its shape
    are {now} required to improve its size, hence
    volume, estimate.

%  pr=[1., 0.27/0.12]                            
%  d=940.                                        ;-Brown
%  d=850.                                        ;-Lim
%  m=641.0e18
%  orc= sqrt( d*d/( 1.+pr*10.^(-0.4*2.4) )  )   
%  van=sqrt(d*d-orc*orc)                        
%  print,  orc, van,  (van/orc)^3., (orc/van)^3.
%
%  orc*=1000.
%  van*=1000.
%  d*=1000.
%  print, (6*m/(!PI*d^3.))*( (1+(van/orc)^2.)^1.5 )/(1+(van/orc)^3.) 

%%%%%%%%%%%%%%%%%%%%%%%%%%%%%%%%%%%%%%%%%%%%%%%%%%%%%%%%%%%%%%%%%
%%%%%%%%%%%%%%%%%%%%%%%%%%%%%%%%%%%%%%%%%%%%%%%%%%%%%%%%%%%%%%%%%
%%%%%%%%%%%%%%%%%%%%%%%%%%%%%%%%%%%%%%%%%%%%%%%%%%%%%%%%%%%%%%%%%
\begin{table}
\caption{Orbital elements and physical parameters for the Orcus/Vanth
  system.
  {Uncertainties are the formal 1\,$\sigma$ deviation
    (from Fig.~\ref{fig: aeip}) and do not account for possible
    systematics (that arise from model incompleteness, like, \eg,
    neglecting Orcus J$_2$ which is unknown).} 
\label{tab: orbit}
} 
\centering
\begin{tabular}{c@{ }ccll}
\hline
\hline
Qty. & Value & Uncert. & Description \\
\hline
 $a$      & 9\,030     &  89        & km      & Semi-major axis \\
 $e$      & 0.007      &  0.003     & --      & Eccentricity \\
 $i$      & ($\pm$)21  &  2         & deg.    & Inclination$^1$ \\
 $\Omega$ & 144        & 63         & deg.    & Ascending node$^1$\\
% $\omega$ & -- \\
% $n$      & 0.658\,548 & 0.000\,006 & rad/day & Mean motion \\
 $n$      & 37.732 & 0.002 & deg/day & Mean motion \\
\hline
 $P$ & 9.540\,6         & 0.000\,4 & day & Orbital period \\
 P$_{\rm A}$ & (321, -2)  &     3    & deg.& Pole A coordinates$^2$ \\
 P$_{\rm B}$ & (340, +38) & 3        & deg.& Pole B coordinates$^2$ \\
\hline
 $\mathcal{M}$ &  641         &  19 & $10^{18}$\,kg & Total mass \\
 $\mathcal{D}$ &  850         &  90 & km & Diameter$^3$\\
 $D_{\rm o}$    &  807 or 761  & 100 & km & Orcus diameter$^4$ \\
 $D_{\rm v}$    &  267 or 378  & 100 & km & Vanth diameter$^4$ \\
 $\rho$        & 2250 or 2470 & 885 & kg\,m$^{-3}$ & Bulk density\\
\hline
\end{tabular}
\begin{list}{}{}
  \item[$^1$] In the tangent plane.
  \item[$^2$] Pole coordinates are given in the ecliptic J2000
  reference frame.
  \item[$^3$] Volume-equivalent from radiometry \citep[see][]{2010-AA-518-Lim}.
  \item[$^4$] Assuming alternatively $p_{\rm v}$\,=\,$p_{\rm o}$, or
    $p_{\rm o}$\,=\,0.27 and $p_{\rm v}$\,=\,0.12.
\end{list}
\end{table}
%%%%%%%%%%%%%%%%%%%%%%%%%%%%%%%%%%%%%%%%%%%%%%%%%%%%%%%%%%%%%%%%%
%%%%%%%%%%%%%%%%%%%%%%%%%%%%%%%%%%%%%%%%%%%%%%%%%%%%%%%%%%%%%%%%%
%%%%%%%%%%%%%%%%%%%%%%%%%%%%%%%%%%%%%%%%%%%%%%%%%%%%%%%%%%%%%%%%%

  \subsection{Formation Mechanism\label{ssec: formation}}
    \indent Trans-Neptunian binaries can be used to impose constraints
    on various models of Solar System formation evolution, \eg, to
    constrain Neptune's migration history
    \citep{2011-ApJ-730-Murray-Clay}.
    Several formation mechanisms have been proposed and each
    leads to different predictions for the physical and orbital
    properties of TNBs. These formation mechanisms can, broadly, be
    broken down into four classes: collision
    \citep{2002-Icarus-160-Weidenschilling},
    rotational fission
    \citep{2011-AA-525-Ortiz},
    capture
    \citep[\eg,][]{2002-Nature-420-Goldreich,
      2004-Nature-427-Funato, 
      2005-MNRAS-360-Astakhov, 
      2007-MNRAS-379-Lee, 
      2010-CeMDA-106-Gamboa-Suarez}, and gravitational collapse
    \citep{2010-AJ-140-Nesvorny}. \\
    \indent In the collision model of
    \citet{2002-Icarus-160-Weidenschilling} two
    objects collide inside the Hill sphere of a third object. These
    objects then fuse into a single object thereby producing a
    binary. The capture models of
    \citet{2002-Nature-420-Goldreich} rely on two
    objects interpenetrating their mutual Hill sphere and then being
    stabilized either through dynamical friction (the L$^2$ mechanism)
    or through a scattering event with a third, similarly sized,
    object (the L$^3$ mechanism).
    \citet{2004-Nature-427-Funato} proposed a
    hybrid collision-capture mechanism. Initially two objects collide
    to produce a binary whose components (as is usual for a collision)
    have quite different masses. Subsequently, exchange ``reactions''
    with larger third bodies displace the secondary and so ramp up the
    mass ratio. This eventually leads to binaries having similarly
    sized partners. However, this mechanism appears to lead to orbital
    properties (in particular, ellipticities) dissimilar to those
    actually observed
    \citep[\eg,][]{2003-EMP-92-Noll, 2005-MNRAS-360-Astakhov}.  \\
    \indent A further capture scenario, chaos-assisted capture
    \citep[CAC, see][]{2005-MNRAS-360-Astakhov, 2007-MNRAS-379-Lee},
    was originally proposed
    to explain the capture of irregular moons at the giant planets
    \citep[\eg,][]{2003-Nature-423-Astakhov}.
    The scenario for the
    formation of TNBs is as follows: two objects initially become
    caught up in very long living, yet ultimately unstable, chaotic
    orbits within their mutual Hill sphere. During this phase the
    binary may be permanently captured and subsequently hardened
    through multiple scattering encounters with relatively small
    ``intruder'' bodies. The CAC model predicts similarly-sized binary
    components whose mutual orbits are in good agreement with
    observations. In fact, it was proposed that the propensity to form
    similar sized binary partners is a direct fingerprint of chaos
    assisted capture. A further specific prediction of the CAC
    mechanism concerns the distribution of retrograde to prograde
    mutual orbits, \ie, mutual orbit inclinations
    \citep{2005-MNRAS-360-Astakhov, 2007-MNRAS-379-Lee}.
    In particular, retrograde mutual
    orbits are predicted to be relatively common. The CAC mechanism
    has been criticized by
    \citet{2008-ApJ-673-Schlichting}
    who argue, \eg, that formation via transient, chaotic binaries is
    not as important as the L$^2$ and L$^3$ mechanisms.
    These objections are addressed
    in some detail by \citet{2010-CeMDA-106-Gamboa-Suarez}. \\
    \indent Finally, \citet{2010-AJ-140-Nesvorny} recently proposed a
    model in which TNBs formed during gravitational collapse. Angular
    momentum considerations in the planetesimal disk explain the
    formation of binaries rather than condensation into a single
    object. This gravitational instability model predicts identical
    compositions and colors for TNB partners and also inclinations
    generally $i \le 50^\circ$, \ie, retrograde mutual orbits are
    predicted to be rare.
    Unfortunately 
    true inclinations of mutual TNB orbits is currently
    available {for a handful of systems only
      \citep[see][]{2011-Icarus-213-Grundy}.}
    For example, Table~\ref{tab: orbit} and
    Figs.~\ref{fig: pred} and~\ref{fig: aeip} show
    that only knowledge of the true inclination of the mutual orbit
    (\ie, by making further observations to remove the symmetry
    ambiguity) will reveal if the mutual orbit of the Orcus/Vanth
    system is prograde or retrograde. \\
    \indent We remark that the CAC model also allows for the direct
    formation of almost circular orbits similar to that of
    Orcus/Vanth. This mechanism was proposed as a possible origin of
    the apparently almost circular orbit of the TNB 2001 QW$_{322}$
    \citep{2008-Science-322-Petit, 2010-CeMDA-106-Gamboa-Suarez}.
    It is conceivable that the Orcus/Vanth system was produced in a
    similar manner. 
    In summary, determining the colors and compositions, as well as
    the mutual orbital true inclinations of TNBs
    \citep{2008-ApJ-673-Schlichting}
    remain important quests. Knowledge of these properties will
    provide important information to constrain, and to distinguish
    between, the variously proposed TNB formation models.

  \subsection{Stellar occultation\label{sec: occult}}
    \indent Observations of stellar occultation phenomena are of high
    interest to derive, in a direct and most precise way, the size of
    the occulting body. When several chords are gathered one can
    derive the projected shape contour
    \citep[see][for a extensive summary]{1989-AsteroidsII-Millis}. 
    Moreover such observations can
    reveal presence of a teneous atmosphere, and then derive its pressure
    profile
    \citep[\eg,][]{2003-Nature-424-Sicardy}. \\
    \indent In the case of a binary system it is also important to be
    able to predict the path of each component more than the one of
    the centre of gravity \citep{2010-AA-515-Assafin}.
    Given the large separation of Vanth with
    respect to Orcus ($\sim$9\,000\,km),
    the prediction of an event by the secondary can be either
    deplaced in time by several minutes when the binary system is
    oriented in the direction of apparent motion, or fully decoupled
    from the one of the primary otherwise. Indeed, the occultation
    path by the primary may not be visible on Earth while the one by
    the secondary could be visible.

    \indent Owing to the current uncertainty on TNOs ephemerides,
    precise prediction of occultation requires ``last minute''
    astrometry measurements with respect to background star
    to obtain reliable predictions of the occultation track
    on Earth. 
    Besides, the relative position of the secondary is needed
    to a precision of approximately 100\,km
    to enable putting in place
    useful observation campaigns.
    Given the improvement on the period estimate,
    and the orbital phase reset provided by
    our astrometric point obtained in 2010, we can 
    derive better prediction for the location of the secondary and
    subsequently better prediction for stellar occultations by Vanth. 
    Given the possible mass-ratio range the path on the Earth
    of the primary can also be shifted from the position of the
    centre-of-mass.
    Assuming the components have
    same albedo the mass ratio is of the order of
    $\approx$$0.02^{^{+0.07}_{-0.02}}$, corresponding to a shift of
    $180^{^{+630}_{-180}}\,$km.  
    We list in Table~\ref{tab: occ} the relative position of Vanth
    around Orcus for upcoming stellar occultations.
    {The relative positions in Table~\ref{tab: occ} corresponds to the
      nominal values; the associated error distribution is not exactly
      Gaussian, but an ellipse curved and stretched along the
      trajectory (similar to Fig.~\ref{fig: pred}).
      The one sigma uncertainty is
      large, approximately $\pm$450\,km along the secondary's nominal
      trajectory (about $\pm$120\,km across), for both occultations.
      Depending on the occultation occurence (motion relative
      to the star and orientation of this uncertainty ellipse) it can
      correspond to a shift of the occultation path on Earth (or miss)
      or a shift in time. }
    Updates are available
    online\footnote{\href{http://www.lesia.obspm.fr/perso/bruno-sicardy/}{http://www.lesia.obspm.fr/perso/bruno-sicardy/}}.

%%%%%%%%%%%%%%%%%%%%%%%%%%%%%%%%%%%%%%%%%%%%%%%%%%%%%%%%%%%%%%%%%
%%%%%%%%%%%%%%%%%%%%%%%%%%%%%%%%%%%%%%%%%%%%%%%%%%%%%%%%%%%%%%%%%
%%%%%%%%%%%%%%%%%%%%%%%%%%%%%%%%%%%%%%%%%%%%%%%%%%%%%%%%%%%%%%%%%
\begin{table}
\caption{Relative positions ($\delta$x, $\delta$y)
  of Vanth with respect to Orcus 
  for future occultation events
  ($\delta$x positive through East).
  {Uncertainty on position draws an ellipse of
    $\pm$\,450\,$\times$\,120 on Earth, around the reported positions
    (see text for details).}
  \label{tab: occ}
} 
\centering
\begin{tabular}{ccc}
\hline
\hline
Date & $\delta$x (km) & $\delta$y (km) \\
\hline
  2012 Mar 21 & -6977 & -5259 \\
  2012 Nov 16 & -8937 &   347 \\
\hline
\end{tabular}
\end{table}
%%%%%%%%%%%%%%%%%%%%%%%%%%%%%%%%%%%%%%%%%%%%%%%%%%%%%%%%%%%%%%%%%
%%%%%%%%%%%%%%%%%%%%%%%%%%%%%%%%%%%%%%%%%%%%%%%%%%%%%%%%%%%%%%%%%
%%%%%%%%%%%%%%%%%%%%%%%%%%%%%%%%%%%%%%%%%%%%%%%%%%%%%%%%%%%%%%%%%

%------------------------------------------------%
%------------------------------------------------%
%---- TAG ---------------------------------------%
%------------------------------------------------%
%------------------------------------------------%
\section{Conclusion}

  From spectro-imaging measurements of the Orcus-Vanth system we
  were able to obtain a spectrum of Orcus, H and K magnitude for Vanth,
  and determine their mutual position.
  We searched for possible weak
  bands in the spectrum of Orcus,
  limiting the possible amount of CH$_4$ to no more than $\sim$2\%,
  and ethane to $\sim$5\%.
  The presence of other compound(s) is required to 
  explain the depth of the absorption band detected at 2.2\,\micron.
  Presence of hydrated ammonia, or ammonium,
%  up to a couple of percent
  could explain this
  depth, but the lack of distinguishable features for these species
  in the
  observed wavelength range forbid any strong conclusion to be
  drawn.\\
  \indent {Vanth appears}
  slightly redder than
  Orcus in the visible to near-infrared wavelength range,
  although uncertainties in the measurements are large.
  While this suggests that Vanth's
  composition, or age, is different from Orcus,
  it does not constrain formation scenarios
  (collision, co-formation, or capture).  
  By detecting Vanth at a new position three years after previous orbit
  calculations by \citet{2010-AJ-139-Brown}
  we are able to reset the phase of the orbit, allowing prediction of
  Vanth's track on Earth during future stellar occultation events.

%------------------------------------------------%
%------------------------------------------------%
%---- TAG ---------------------------------------%
%------------------------------------------------%
%------------------------------------------------%
\section{Acknowledgments}

  We thank the staff of ESO's Paranal
  observatory for their assistance in obtaining this data.
  This research used IMCCE's Miriade
  \citep{2008-ACM-Berthier} VO tool, and 
  NASA's Astrophysics Data System.
  We acknowledge support from the Faculty of the European Space
  Astronomy Centre (ESAC) for FD visit.
  PL is grateful for financial support from a Michael West Fellowship
  and from the Royal Society in the form of a Newton Fellowship.
  {We thank our anonymous referee for his constructive comments.}
%
%  It is a pleasure to acknowledge
%  Erwan ``Guantanamo survivor'' Treguier for his
%  entertainment.

%%%--- Bibliography ---%%%
%  \bibliographystyle{aa}
%  \bibliography{biblio}

\begin{thebibliography}{59}
\expandafter\ifx\csname natexlab\endcsname\relax\def\natexlab#1{#1}\fi

\bibitem[{Assafin {et~al.}(2010)Assafin, Camargo, Vieira~Martins, Andrei,
  Sicardy, Young, da~Silva~Neto, \& Braga-Ribas}]{2010-AA-515-Assafin}
Assafin, M., Camargo, J.~I.~B., Vieira~Martins, R., {et~al.} 2010, Astronomy
  and Astrophysics, 515, A32

\bibitem[{Astakhov {et~al.}(2003)Astakhov, Burbanks, Wiggins, \&
  Farrelly}]{2003-Nature-423-Astakhov}
Astakhov, S.~A., Burbanks, A.~D., Wiggins, S., \& Farrelly, D. 2003, Nature,
  423, 264

\bibitem[{Astakhov {et~al.}(2005)Astakhov, Lee, \&
  Farrelly}]{2005-MNRAS-360-Astakhov}
Astakhov, S.~A., Lee, E.~A., \& Farrelly, D. 2005, Monthly Notices of the Royal
  Astronomical Society, 360, 401

\bibitem[{Barucci {et~al.}(2008{\natexlab{a}})Barucci, Brown, Emery, \&
  Merlin}]{2008-SSBN-3-Barucci}
Barucci, M.~A., Brown, M.~E., Emery, J.~P., \& Merlin, F. 2008{\natexlab{a}},
  The Solar System Beyond Neptune, 143

\bibitem[{Barucci {et~al.}(2008{\natexlab{b}})Barucci, Merlin, Guilbert,
  de~Bergh, Alvarez-Candal, Hainaut, Doressoundiram, Dumas, Owen, \&
  Coradini}]{2008-AA-479-Barucci}
Barucci, M.~A., Merlin, F., Guilbert, A., {et~al.} 2008{\natexlab{b}},
  Astronomy and Astrophysics, 479, L13

\bibitem[{Berthier {et~al.}(2008)Berthier, Hestroffer, Carry, {\v D}urech,
  Tanga, Delbo, \& Vachier}]{2008-ACM-Berthier}
Berthier, J., Hestroffer, D., Carry, B., {et~al.} 2008, LPI Contributions,
  1405, 8374

\bibitem[{Berthier \& Marchis(2001)}]{2001-DPS-33-Berthier}
Berthier, J. \& Marchis, F. 2001, in Bulletin of the American Astronomical
  Society, Vol.~33, 1049

\bibitem[{Bonnet {et~al.}(2004)Bonnet, Abuter, Baker, Bornemann, Brown,
  Castillo, Conzelmann, Damster, Davies, Delabre, Donaldson, Dumas, Eisenhauer,
  Elswijk, Fedrigo, Finger, Gemperlein, Genzel, Gilbert, Gillet, Goldbrunner,
  Horrobin, Ter~Horst, Huber, Hubin, Iserlohe, Kaufer, Kissler-Patig, Kragt,
  Kroes, Lehnert, Lieb, Liske, Lizon, Lutz, Modigliani, Monnet, Nesvadba,
  Patig, Pragt, Reunanen, R{\"o}hrle, Rossi, Schmutzer, Schoenmaker, Schreiber,
  Stroebele, Szeifert, Tacconi, Tecza, Thatte, Tordo, van~der Werf, \&
  Weisz}]{2004-Msngr-117-Bonnet}
Bonnet, H., Abuter, R., Baker, A., {et~al.} 2004, The Messenger, 117, 17

\bibitem[{Britt {et~al.}(2002)Britt, Yeomans, Housen, \&
  Consolmagno}]{2002-AsteroidsIII-4.2-Britt}
Britt, D.~T., Yeomans, D.~K., Housen, K.~R., \& Consolmagno, G.~J. 2002,
  Asteroids III, 485

\bibitem[{Brown {et~al.}(2010)Brown, Ragozzine, Stansberry, \&
  Fraser}]{2010-AJ-139-Brown}
Brown, M.~E., Ragozzine, D., Stansberry, J., \& Fraser, W.~C. 2010,
  Astronomical Journal, 139, 2700

\bibitem[{Brown {et~al.}(2006)Brown, van Dam, Bouchez, Le~Mignant, Campbell,
  Chin, Conrad, Hartman, Johansson, Lafon, Rabinowitz, Stomski~Jr., Summers,
  Trujillo, \& Wizinowich}]{2006-ApJ-639-Brown}
Brown, M.~E., van Dam, M.~A., Bouchez, A.~H., {et~al.} 2006, Astrophysical
  Journal, 639, 43–46

\bibitem[{Carry {et~al.}(2010)Carry, Vernazza, Dumas, \&
  Fulchignoni}]{2010-Icarus-205-Carry-b}
Carry, B., Vernazza, P., Dumas, C., \& Fulchignoni, M. 2010, Icarus, 205, 473

\bibitem[{Cook {et~al.}(2007)Cook, Desch, Roush, Trujillo, \&
  Geballe}]{2007-AJ-663-Cook}
Cook, J.~C., Desch, S.~J., Roush, T.~L., Trujillo, C.~A., \& Geballe, T.~R.
  2007, Astrophysical Journal, 663, 1406

\bibitem[{Cooper {et~al.}(2003)Cooper, Christian, Richardson, \&
  Wang}]{2003-EMP-92-Cooper}
Cooper, J.~F., Christian, E.~R., Richardson, J.~D., \& Wang, C. 2003, Earth
  Moon and Planets, 92, 261

\bibitem[{Cottin {et~al.}(2003)Cottin, Moore, \&
  B{\'e}nilan}]{2003-ApJ-590-Cottin}
Cottin, H., Moore, M.~H., \& B{\'e}nilan, Y. 2003, Astrophysical Journal, 590,
  874

\bibitem[{de~Bergh {et~al.}(2005)de~Bergh, Delsanti, Tozzi, Dotto,
  Doressoundiram, \& Barucci}]{2005-AA-437-deBergh}
de~Bergh, C., Delsanti, A., Tozzi, G.~P., {et~al.} 2005, Astronomy and
  Astrophysics, 437, 1115

\bibitem[{Delsanti {et~al.}(2010)Delsanti, Merlin, Guilbert-Lepoutre, Bauer,
  Yang, \& Meech}]{2010-AA-520-Delsanti}
Delsanti, A., Merlin, F., Guilbert-Lepoutre, A., {et~al.} 2010, Astronomy and
  Astrophysics, 520, A40

\bibitem[{DeMeo {et~al.}(2010)DeMeo, Barucci, Merlin, Guilbert-Lepoutre,
  Alvarez-Candal, Delsanti, Fornasier, \& de~Bergh}]{2010-AA-521-DeMeo}
DeMeo, F.~E., Barucci, M.~A., Merlin, F., {et~al.} 2010, Astronomy and
  Astrophysics, 521, A35

\bibitem[{DeMeo {et~al.}(2009)DeMeo, Fornasier, Barucci, Perna, Protopapa,
  Alvarez-Candal, Delsanti, Doressoundiram, Merlin, \&
  de~Bergh}]{2009-AA-493-DeMeo}
DeMeo, F.~E., Fornasier, S., Barucci, M.~A., {et~al.} 2009, Astronomy and
  Astrophysics, 493, 283

\bibitem[{Dumas {et~al.}(2011)Dumas, Carry, Hestroffer, \&
  Merlin}]{2011-AA-528-Dumas}
Dumas, C., Carry, B., Hestroffer, D., \& Merlin, F. 2011, Astronomy and
  Astrophysics, 528, A105

\bibitem[{Eisenhauer {et~al.}(2003)Eisenhauer, Abuter, Bickert,
  Biancat-Marchet, Bonnet, Brynnel, Conzelmann, Delabre, Donaldson, Farinato,
  Fedrigo, Genzel, Hubin, Iserlohe, Kasper, Kissler-Patig, Monnet, Roehrle,
  Schreiber, Stroebele, Tecza, Thatte, \& Weisz}]{2003-SPIE-1548-Eisenhauer}
Eisenhauer, F., Abuter, R., Bickert, K., {et~al.} 2003, SPIE, 4841, 1548

\bibitem[{Ferraz-Mello {et~al.}(2008)Ferraz-Mello, Rodr{\'{\i}}guez, \&
  Hussmann}]{2008-CeMDA-101-Ferraz-Mello}
Ferraz-Mello, S., Rodr{\'{\i}}guez, A., \& Hussmann, H. 2008, Celestial
  Mechanics and Dynamical Astronomy, 101, 171

\bibitem[{Fornasier {et~al.}(2004)Fornasier, Dotto, Barucci, \&
  Barbieri}]{2004-AA-422-Fornasier}
Fornasier, S., Dotto, E., Barucci, M.~A., \& Barbieri, C. 2004, Astronomy and
  Astrophysics, 422, L43

\bibitem[{Funato {et~al.}(2004)Funato, Makino, Hut, Kokubo, \&
  Kinoshita}]{2004-Nature-427-Funato}
Funato, Y., Makino, J., Hut, P., Kokubo, E., \& Kinoshita, D. 2004, Nature,
  427, 518

\bibitem[{Gamboa~Su{\'a}rez {et~al.}(2010)Gamboa~Su{\'a}rez, Hestroffer, \&
  Farrelly}]{2010-CeMDA-106-Gamboa-Suarez}
Gamboa~Su{\'a}rez, A., Hestroffer, D., \& Farrelly, D. 2010, Celestial
  Mechanics and Dynamical Astronomy, 106, 245

\bibitem[{Goldreich {et~al.}(2002)Goldreich, Lithwick, \&
  Sari}]{2002-Nature-420-Goldreich}
Goldreich, P., Lithwick, Y., \& Sari, R. 2002, Nature, 420, 643

\bibitem[{Grundy {et~al.}(2011)Grundy, Noll, Nimmo, Roe, Buie, Porter,
  Benecchi, Stephens, Levison, \& Stansberry}]{2011-Icarus-213-Grundy}
Grundy, W.~M., Noll, K.~S., Nimmo, F., {et~al.} 2011, Icarus, 213, 678

\bibitem[{Grundy \& Schmitt(1998)}]{1998-JGR-103-Grundy}
Grundy, W.~M. \& Schmitt, B. 1998, Journal of Geophysical Research, 103, 25809

\bibitem[{Guilbert {et~al.}(2009)Guilbert, Alvarez-Candal, Merlin, Barucci,
  Dumas, de~Bergh, \& Delsanti}]{2009-Icarus-201-Guilbert}
Guilbert, A., Alvarez-Candal, A., Merlin, F., {et~al.} 2009, Icarus, 201, 272

\bibitem[{Hapke(1993)}]{hapke-theory}
Hapke, B. 1993, {Theory of reflectance and emittance spectroscopy} (Cambridge
  University Press)

\bibitem[{Hestroffer {et~al.}(2005)Hestroffer, Vachier, \&
  Balat}]{2005-EMP-97-Hestroffer}
Hestroffer, D., Vachier, F., \& Balat, B. 2005, Earth Moon and Planets, 97, 245

\bibitem[{Hilton(2002)}]{2002-AsteroidsIII-2.2-Hilton}
Hilton, J.~L. 2002, Asteroids III, 103

\bibitem[{Khare {et~al.}(1984)Khare, Sagan, Arakawa, Suits, Callcott, \&
  Williams}]{1984-Icarus-60-Khare}
Khare, B.~N., Sagan, C., Arakawa, E.~T., {et~al.} 1984, Icarus, 60, 127

\bibitem[{Khare {et~al.}(1993)Khare, Thompson, Cheng, Chyba, Sagan, Arakawa,
  Meisse, \& Tuminello}]{1993-Icarus-103-Khare}
Khare, B.~N., Thompson, W.~R., Cheng, L., {et~al.} 1993, Icarus, 103, 290

\bibitem[{Kozai(1962)}]{1962-AJ-67-Kozai}
Kozai, Y. 1962, Astronomical Journal, 67, 591

\bibitem[{Lee {et~al.}(2007)Lee, Astakhov, \& Farrelly}]{2007-MNRAS-379-Lee}
Lee, E.~A., Astakhov, S.~A., \& Farrelly, D. 2007, Monthly Notices of the Royal
  Astronomical Society, 379, 229

\bibitem[{Levi \& Podolak(2009)}]{2009-Icarus-202-Levi}
Levi, A. \& Podolak, M. 2009, Icarus, 202, 681

\bibitem[{Lim {et~al.}(2010)Lim, Stansberry, M{\"u}ller, Mueller, Lellouch,
  Kiss, Santos-Sanz, Vilenius, Protopapa, Moreno, Delsanti, Duffard, Fornasier,
  Groussin, Harris, Henry, Horner, Lacerda, Mommert, Ortiz, Rengel, Thirouin,
  Trilling, Barucci, Crovisier, Doressoundiram, Dotto,
  Guti{\'e}rrez~Buenestado, Hainaut, Hartogh, Hestroffer, Kidger, Lara,
  Swinyard, \& Thomas}]{2010-AA-518-Lim}
Lim, T.~L., Stansberry, J., M{\"u}ller, T.~G., {et~al.} 2010, Astronomy and
  Astrophysics, 518, L148

\bibitem[{Markwardt(2009)}]{2009-ASPC-411-Markwardt}
Markwardt, C.~B. 2009, in Astronomical Society of the Pacific Conference
  Series, ed. {D.~A.~Bohlender, D.~Durand, \& P.~Dowler}, Vol. 411, 251

\bibitem[{Merline {et~al.}(2002)Merline, Weidenschilling, Durda, Margot,
  Pravec, \& Storrs}]{2002-AsteroidsIII-2.2-Merline}
Merline, W.~J., Weidenschilling, S.~J., Durda, D.~D., {et~al.} 2002, Asteroids
  III, 289

\bibitem[{Millis \& Dunham(1989)}]{1989-AsteroidsII-Millis}
Millis, R.~L. \& Dunham, D.~W. 1989, Asteroids II, 148

\bibitem[{Modigliani {et~al.}(2007)Modigliani, Hummel, Abuter, Amico, Balleste,
  Davies, Dumas, Horrobin, Neeser, Kissler-Patig, Peron, Rehunanen, Schreiber,
  \& Szeifert}]{2007-arXiv-Modigliani}
Modigliani, A., Hummel, W., Abuter, R., {et~al.} 2007, ArXiv Astrophysics
  e-prints

\bibitem[{M{\"u}ller {et~al.}(2009)M{\"u}ller, Lellouch, B{\"o}hnhardt,
  Stansberry, Barucci, Crovisier, Delsanti, Doressoundiram, Dotto, Duffard,
  Fornasier, Groussin, Guti{\'e}rrez, Hainaut, Harris, Hartogh, Hestroffer,
  Horner, Jewitt, Kidger, Kiss, Lacerda, Lara, Lim, Mueller, Moreno, Ortiz,
  Rengel, Santos-Sanz, Swinyard, Thomas, Thirouin, \&
  Trilling}]{2009-EMP-105-Muller}
M{\"u}ller, T.~G., Lellouch, E., B{\"o}hnhardt, H., {et~al.} 2009, Earth Moon
  and Planets, 105, 209

\bibitem[{Murray-Clay \& Schlichting(2011)}]{2011-ApJ-730-Murray-Clay}
Murray-Clay, R.~A. \& Schlichting, H.~E. 2011, Astrophysical Journal, 730, 132

\bibitem[{Nesvorn{\'y} {et~al.}(2010)Nesvorn{\'y}, Youdin, \&
  Richardson}]{2010-AJ-140-Nesvorny}
Nesvorn{\'y}, D., Youdin, A.~N., \& Richardson, D.~C. 2010, Astronomical
  Journal, 140, 785

\bibitem[{Noll(2003)}]{2003-EMP-92-Noll}
Noll, K.~S. 2003, Earth Moon and Planets, 92, 395

\bibitem[{Ortiz {et~al.}(2011)Ortiz, Cikota, Cikota, Hestroffer, Thirouin,
  Morales, Duffard, Gil-Hutton, Santos-Sanz, \&
  de~La~Cueva}]{2011-AA-525-Ortiz}
Ortiz, J.~L., Cikota, A., Cikota, S., {et~al.} 2011, Astronomy and
  Astrophysics, 525, A31

\bibitem[{Petit {et~al.}(2008)Petit, Kavelaars, Gladman, Margot, Nicholson,
  Jones, Parker, Ashby, Campo~Bagatin, Benavidez, Coffey, Rousselot, Mousis, \&
  Taylor}]{2008-Science-322-Petit}
Petit, J.-M., Kavelaars, J.~J., Gladman, B.~J., {et~al.} 2008, Science, 322,
  432

\bibitem[{Quirico \& Schmitt(1997)}]{1997-Icarus-127-Quirico}
Quirico, E. \& Schmitt, B. 1997, Icarus, 127, 354

\bibitem[{Schaller \& Brown(2007)}]{2007-ApJ-659-Schaller}
Schaller, E.~L. \& Brown, M.~E. 2007, Astrophysical Journal, 659, L61

\bibitem[{Schlichting \& Sari(2008)}]{2008-ApJ-673-Schlichting}
Schlichting, H.~E. \& Sari, R. 2008, Astrophysical Journal, 673, 1218

\bibitem[{Schmitt {et~al.}(1998)Schmitt, Quirico, Trotta, \&
  Grundy}]{1998-ASSL-227-Schmitt}
Schmitt, B., Quirico, E., Trotta, F., \& Grundy, W.~M. 1998, in Astrophysics
  and Space Science Library, Vol. 227, Solar System Ices, ed. {B.~Schmitt,
  C.~de Bergh, \& M.~Festou}, 199--+

\bibitem[{Sicardy {et~al.}(2003)Sicardy, Widemann, Lellouch, Veillet,
  Cuillandre, Colas, Roques, Beisker, Kretlow, Lagrange, Gendron, Lacombe,
  Lecacheux, Birnbaum, Fienga, Leyrat, Maury, Raynaud, Renner, Schultheis,
  Brooks, Delsanti, Hainaut, Gilmozzi, Lidman, Spyromilio, Rapaport,
  Rosenzweig, Naranjo, Porras, D{\'{\i}}az, Calder{\'o}n, Carrillo, Carvajal,
  Recalde, Cavero, Montalvo, Barr{\'{\i}}a, Campos, Duffard, \&
  Levato}]{2003-Nature-424-Sicardy}
Sicardy, B., Widemann, T., Lellouch, E., {et~al.} 2003, Nature, 424, 168

\bibitem[{Stansberry {et~al.}(2008)Stansberry, Grundy, Brown, Cruikshank,
  Spencer, Trilling, \& Margot}]{2008-SSBN-3-Stansberry}
Stansberry, J., Grundy, W., Brown, M.~E., {et~al.} 2008, The Solar System
  Beyond Neptune, 161

\bibitem[{Strazzulla \& Palumbo(1998)}]{1998-PSS-46-Strazzulla}
Strazzulla, G. \& Palumbo, M.~E. 1998, Planetary and Space Science, 46, 1339

\bibitem[{Thirouin {et~al.}(2010)Thirouin, Ortiz, Duffard, Santos-Sanz,
  Aceituno, \& Morales}]{2010-AA-522-Thirouin}
Thirouin, A., Ortiz, J.~L., Duffard, R., {et~al.} 2010, Astronomy and
  Astrophysics, 522, A93

\bibitem[{Trujillo {et~al.}(2005)Trujillo, Brown, Rabinowitz, \&
  Geballe}]{2005-ApJ-627-Trujillo}
Trujillo, C.~A., Brown, M.~E., Rabinowitz, D.~L., \& Geballe, T.~R. 2005,
  Astrophysical Journal, 627, 1057

\bibitem[{Tryka {et~al.}(1994)Tryka, Brown, Chruikshank, Owen, Geballe, \&
  de~Bergh}]{1994-Icarus-112-Tryka}
Tryka, K.~A., Brown, R.~H., Chruikshank, D.~P., {et~al.} 1994, Icarus, 112, 513

\bibitem[{Weidenschilling(2002)}]{2002-Icarus-160-Weidenschilling}
Weidenschilling, S.~J. 2002, Icarus, 160, 212

\end{thebibliography}

\end{document}